\documentclass{mn2e}
\usepackage{psfig}

\def\ltsima{$\; \buildrel < \over \sim \;$}
\def\lsim{\lower.5ex\hbox{\ltsima}}
\def\gtsima{$\; \buildrel > \over \sim \;$}
\def\gsim{\lower.5ex\hbox{\gtsima}}

\begin{document} 

\title[carbonaceous dust nucleation]{Thermal fluctuations and
  nanoscale effects in the nucleation of carbonaceous dust grains}

\author[Keith \& Lazzati]
{Adam C. Keith, Davide Lazzati \\ 
Department of Physics, North Carolina State University, 2401 Stinson Drive, 
Raleigh, NC 27695-8202}

\maketitle

\begin{abstract}
  We investigate the role of thermal fluctuations and of the finite
  number of monomers in small clusters of carbon atoms on the
  nucleation rate of carbonaceous grains. Thermal fluctuations are due
  to the quantized nature of the energy exchanges between the
  clusters, the gas, and the radiation field. Nanoscale effects modify
  the spontaneous detachment of monomers due to the finite amount of
  internal energy contained in small clusters. We find that both
  corrections have a big impact on the stability of the clusters and
  on the rate of nucleation. We implement our model within a Monte
  Carlo code to derive the new stability conditions for clusters as
  well as nucleation rates. Due to computing limitations, we can
  explore the consequences of this approach only at high temperatures,
  at which particle interactions are not much less frequent than
  photon interactions. We found that the combined effect of the
  detachment correction and the temperature fluctuations produces
  faster nucleation. We also found that the nucleation rate depends on
  the composition of the gas and not only on the partial pressure of
  the compound that condensates into grains. This is a unique result
  of this model that can be used to prove or disprove it.
\end{abstract}


\section{Introduction} 

Dust is one of the main constituents of the interstellar medium (ISM),
making up to 1 per cent of the baryonic mass of galaxies that is not
in stars (Mathis, Rumpl \& Nordsieck 1977). The average composition of
interstellar dust is made of Polycyclic Aromatic Hydrocarbons (PAHs),
carbonaceous grains, and silicates (Weingartner \& Draine 2001), each
of which contains particles ranging from several angstroms to a few
microns in size. Dust plays many important roles in astrophysics: it
provides attenuation of radiation (Cardelli, Clayton \& Mathis 1989),
induces polarization in the optical and UV frequencies (Serkowski,
Mathewson \& Ford 1975), provides a safe environment for molecular
chemistry (Brown 1990), cools protostellar clouds allowing for the
formation of population II.5 stars (Schneider et al. 2006), and,
finally, provides the building blocks for the formation of terrestrial
planets (Lissauer 1993).

Despite such fundamental roles played by dust, the physics of dust
nucleation is still highly debated, and the sites, rates, and
composition of the dust produced in the different environments are
poorly known. Historically, dust nucleation has been treated as an out
of equilibrium perturbation of a thermodynamical process (Becker \&
Doring 1935; Feder et al. 1966; Kashchiev 2000). Under this approach,
the dust/vapor mix is assumed to be in thermodynamical equilibrium and
able to exchange energy in a continuous way. A critical cluster size
can be identified as the cluster with the highest Gibbs free energy
and the nucleation rate is computed as the flow of particles through
the critical cluster size (Becker \& Doring 1935; Feder et al. 1996;
Kashchiev 2000). Even though this ``classical'' theory can give an
explanation to the qualitative features of nucleation (the presence of
a threshold below which nucleation is forbidden, the existence of
supersaturation, and the steep increase of the nucleation rate with
both temperature and saturation), the quantitative comparison between
prediction and rates is still problematic (W\"olk \& Strey 2001). To
make things worse, any attempt to improve the theory has resulted in a
deterioration of the quantitative comparison.

The adoption of the thermodynamical approach for astrophysical dust
nucleation was criticized already by Donn \& Nuth (1985) who pointed
out problems with the time scale for obtaining thermodynamic
equilibrium as well as discrepancies between the theory and
experiments. An alternative to the thermodynamical treatment is the
kinetic treatment of nucleation (Becker \& Doring 1935, Kashchiev
2000).  The kinetic theory derives nucleation rates by analyzing the
accretion and ejection rates of monomers from sub-critical
clusters. The most fundamental difference is that in the kinetic
theory the nucleation rate depends on the properties of grains of all
sizes (and especially of the unstable ones) while in the thermodynamic
theory the nucleation rate depends only on the properties of the
critical cluster. Another advantage of the kinetic theory is that of
allowing the consideration of radiation as a player in the nucleation
game, as both an energy sink (Lazzati 2008) and as an energy source
(this paper). Astrophysical dust nucleation is more complex than lab
nucleation (e.g. of water, W\"olk \& Strey 2001) for at least two
reasons. First, the conditions of nucleation in astrophysics are very
different from lab conditions and the verification of any theory that
predicts nucleation rates of a particular dust compound can be
performed only very indirectly. Second, some of the components of
interstellar dust, such as silicates, do not have a vapor phase and
the nucleation process must include the chemical bonding of the atoms
into the basic monomer of the grain. We call this process chemical
nucleation to distinguish it from homogeneous nucleation (the
nucleation of a vapor on its own liquid or solid phase) and from
heterogeneous nucleation (the nucleation of a compound on a different
substrate). This difficulty of chemical nucleation has been approached
in two ways. Kozasa et al. (1989, 1991, see also Nozawa et al. 2003,
2008, 2010; Todini \& Ferrara 2001; Schneider et al. 2004; Bianchi \&
Schneider 2007) consider the nucleation process as driven by the atom
that accretes with the lowest frequency, assuming all other compounds
to be already present on the grain surface. It is unclear, however,
how the very first molecules are formed and bound together in a
sub-critical cluster. Cherchneff et al. (2000) and Cherchneff \& Dwek
(2009, 2010) consider instead the chemistry of precursor molecules to
bridge the gap between atoms in the gas phase and sub-critical
clusters onto which the monomers can accrete.

In this paper we address another limitation of the thermodynamical
treatment by considering the effect of discreteness in the energy
exchange between the clusters, the gas, and the radiation field. We
focus on the homogeneous nucleation of carbonaceous grains to avoid
the complication of chemical nucleation, and approximate cluster
surfaces as spherical to simplify calculations. As a consequence of
the quantized nature of cluster-radiation and cluster-atom
interactions, the cluster vibrational temperature can grow well above
the equilibrium temperature and/or decrease below it. We develop a
Monte Carlo code to simulate the thermal history and stability of
small clusters and grains and derive nucleation rates. Due to
technical computing limitations we can, at present, address only the
high temperature regime, since at low temperatures the interactions
are overwhelmingly dominated by exchanges of radiation.

This paper is organized as follows: in Section 2 we describe the
theory of kinetic nucleation and obtain equations for the temperature
fluctuations and ejection rates in small clusters, in Section 3 we
describe the results of the Monte Carlo implementation, and in Section
4 we discuss our results.

\section{Non-equilibrium nucleation} 

The kinetic approach to nucleation (Becker \& Doring 1935; Kashchiev
2000) allows for the computation of nucleation rates in any
non-equilibrium condition. The basic requirement is the knowledge of
the accretion and ejection rates of monomers (in our case Carbon
atoms) from a cluster or grain.\footnote{In this paper we will define a
  cluster as an unstable group of monomers, with a size smaller than the
  critical radius. Similarly, we will define a grain as a stable group of
  monomers, larger than the critical radius. However, when the stability 
  is not the focus of a sentence, cluster and grain are used as synonyms.}

The nucleation rate, i.e., the number of new stable grains formed per
unit time and volume, is given by:
\begin{equation}
  J=n_X f_1 \left[ 1+\sum_{j=2}^\infty \left(
      \prod_{k=2}^j\frac{g_i}{f_i}\right)\right]^{-1}
\label{eq:j}
\end{equation}
where $n_X$ is the density of monomers in the gas phase and $f_i$ is
the rate of the reaction
\begin{equation}
  X_{i-1}+X \to X_i
\end{equation}
where $X$ is a monomer and $X_i$ is a cluster that contains
$i$ monomers. The rate of the reaction 
\begin{equation}
X_i \to X_{i-1}+X
\end{equation}
is instead indicated as $g_i$.

In an equilibrium situation, where the cluster and the gas are at the
same temperature, the attachment rate is given by:
\begin{equation}
f_i=k_s n_X (36\pi)^{1/3} \left(\frac{m_X}{\rho_X}\right)^{2/3}
\sqrt{\frac{kT}{2\pi m_X}}\, i^{2/3}
\label{eq:fi}
\end{equation}
where $k_s\le1$ is the sticking coefficient, $n_X$ is the number
density of the monomer in the gas/vapor phase, and $\rho_X$ is the
density of the solid phase. The detachment rate is instead given by:
\begin{eqnarray}
g_i&=&k_s n_{X,{\rm{eq}}}(T) (36\pi)^{1/3} \left(\frac{m_X}{\rho_X}\right)^{2/3}
\sqrt{\frac{kT}{2\pi m_X}}\, i^{2/3} \times \nonumber \\
&&e^{\left(\frac{32\pi}{3}\right)^{1/3}
  \left(\frac{m_X}{\rho_X}\right)^{2/3} \frac{\sigma}{kT} i^{-1/3}}
\label{eq:gi}
\end{eqnarray}
where $n_{X,{\rm{eq}(T)}}$ is the density of the gas phase at
saturation for the temperature $T$ and $\sigma$ is the surface tension
or surface energy of the condensed phase. The extra exponential term
takes into account the increased detachment rate from small cluster due
to the effect of surface tension (or surface energy) that decreases
the effective binding energy of monomers.

While the attachment $f_i$ only depends on the gas properties (density
and temperature), the detachment frequency depends on the cluster
properties and can change substantially if the assumption of thermal
equilibrium is relaxed. Following Lazzati (2008), we assume that if
the grain temperature $T_{\rm{grain}}$ is different from the gas
temperature $T_{\rm{gas}}$, the detachment frequency is modified as
\begin{eqnarray}
g_i&=&k_s n_{X,{\rm{eq}}}(T_{\rm{grain}})  (36\pi)^{1/3} \left(\frac{m_X}{\rho_X}\right)^{2/3}
\sqrt{\frac{kT_{\rm{grain}}}{2\pi m_X}}\, i^{2/3} \times \nonumber \\
&&e^{\left(\frac{32\pi}{3}\right)^{1/3}
  \left(\frac{m_X}{\rho_X}\right)^{2/3} \frac{\sigma}{kT_{\rm{grain}}}
  i^{-1/3}}
\label{eq:gi2}
\end{eqnarray}
i.e., we assume that the detachment rate depends only on the grain
properties and not on its surroundings. Later, we will
describe an additional modification of the $g_i$ equation to take into
account the reduction in detachment rate from very small and/or very cold clusters
(Guhathakurta \& Draine 1989). Once Eq.~\ref{eq:gi2} is established,
the challenge is to evaluate the cluster temperature to be
used. Lazzati (2008) assumed that the input and output of energy can
be considered continuum and found the equilibrium solution, showing
that in most cases the cluster stabilizes at a temperature that is smaller
than that of the gas, due to the increased radiation losses. In this
paper, we improve on that assumption by considering the quantized
nature of the energy exchanges between the cluster, the gas, and the
radiation field. We consider five interactions responsible for
altering the thermal state of the cluster:
\begin{itemize}
\item attachment of a monomer to the cluster
\item detachment of a monomer from the cluster
\item collision between the cluster and an atom/molecule of an inert
  carrier gas
\item absorption of a photon
\item emission of a photon
\end{itemize}
Due to the statistical nature of the occurrence of all the above
processes, we include them in a Monte Carlo code. At each iteration, the
code stochastically selects the event that would occur first.
For that iteration, energy conservation laws are
applied to determine the thermal state of the cluster after the selected
event occurs. The temperature change is computed through the change of
internal energy, $\Delta U$. The cluster size is also modified if an
attachment or detachment event occurs.

\begin{table}
\begin{tabular}{c|l}
Symbol & Meaning \\ \hline
$b$ & Minimum number of quanta to eject a monomer \\
$C_{\rm{abs}}$ & Correction term for Photon Absorption \\
$C_{\rm{emis}}$ & Correction term for Photon Emission \\
$C_q$ & Correction term for Monomer Detachment \\
$d_f$ & Number of degrees of freedom \\
$E_{\rm{Bind}}$ & Binding energy \\
$E_{\rm{Bind, blk}}$ & Binding energy of a bulk monomer\\
$f_i$ & Attachment rate of monomers onto a cluster of size $i$ \\
$g_i$ & Detachment rate of monomers from a cluster of size $i$ \\
$h_i$ & Collision rate of inert gas atoms with a cluster of size $i$ \\
$H/C$ & Hydrogen/Carbon number density ratio \\
$i_{\rm{C}}$ & Critical cluster size \\
$J$ & Nucleation rate \\
$KE$ & Kinetic energy of a monomer or inert gas particle \\
$k_s$ & Sticking coefficient \\
$m_{CG}$ & Mass of an inert gas particle \\
$m_X$ & Mass of the monomer\\
$n_{CG}$ & Number density of carrier gas \\
$n_X$ & Number density of monomers in the gas phase \\
$n_{X,{\rm{eq}}}$ & Number density of monomers in the gas phase\\ 
$ $ & at vapor pressure\\
$P(\nu,i)$ & Modified Blackbody Distribution \\
$q$ & Total number of quanta in a cluster \\
$S$ & Gas saturation \\
$S(i)$ & Surface area of cluster containing $i$ monomers \\
$T_{\rm{gas}}$ & Temperature of the gas \\
$T_{\rm{grain}}$ & Temperature of the grain \\
$T_{\rm{rad}}$ & Temperature of the radiation field \\
$U$ & Internal energy \\
$X$ & Monomer \\
$X_i$ & Cluster containing $i$ monomers \\
$\alpha_i$ & Photon Absorption rate of a cluster of size $i$ \\
$\Delta U_{\rm{Att}}$ & Change of internal energy for monomer attachment \\
$\varepsilon(\nu,i)$ & Emissivity, frequency and size dependent \\
$\gamma_i$ & Photon Emission rate of a cluster of size $i$ \\
$\rho_X$ & Density of the condensed phase \\
$\sigma$ & Surface energy (tension) of the condensed phase \\
$\sigma_{SB}$ & Stefan-Boltzmann constant \\
$\omega_0$ & Frequency of harmonic oscillators in the Einstein
model 
\end{tabular}
\caption{{Glossary of used symbols}
\label{tab:1}}
\end{table}

\subsection{Monomer Attachment}

The rate of attachment of monomers is given by Eq.~\ref{eq:fi}. The
change of internal energy for a monomer attachment has two components:
the kinetic energy gained from the incoming monomer and the binding
energy that is released at the attachment:
\begin{equation}
\Delta U_{\rm{Att}} = KE(T_{\rm{gas}})+E_{\rm{Bind}}(i\to i+1)
\end{equation}
where the kinetic energy $KE(T_{\rm{gas}})$ is randomly drawn from a
Maxwell-Boltzmann distribution at the gas temperature $T_{\rm{gas}}$,
and $E_{\rm{Bind}}(i)=iE_{Bind,blk} - \sigma S(i)$ is the total
binding energy released in the assembly of $i$ monomers into a cluster
with surface $S(i)$ and binding energy $E_{Bind,blk}$ for a monomer in
the bulk. Thus, the binding energy for an attachment is computed as:
\begin{eqnarray}
  &&E_{\rm{Bind}}(i\to i+1) = \nonumber \\
  &&\left[(i+1)E_{\rm{Bind,blk}} - \sigma S(i+1)\right] - \left[iE_{\rm{Bind,blk}} - \sigma S(i)\right] = \nonumber \\
  &&E_{\rm{Bind,blk}}-\sigma\left[S(i+1)-S(i)\right] = \nonumber \\
  &&E_{\rm{Bind,blk}} - \sigma(36\pi)^{1/3}\left(\frac{m_X}{\rho_X}\right)^{2/3} \left[(i+1)^{2/3}-i^{2/3}\right]
\end{eqnarray}
where we have used the relation $S(i)=(36\pi)^{1/3} \left(\frac{m_Xi}{\rho_X}\right)^{2/3}$.

\subsection{Monomer Detachment}

Similarly, the change of internal energy for a monomer detachment 
has two components: the kinetic energy lost from the outgoing monomer
and the binding energy required to separate the monomer from the
cluster:
\begin{equation}
\Delta U_{\rm{Det}} = -KE(T_{\rm{grain}})+E_{\rm{Bind}}(i\to i-1)
\end{equation}
where the kinetic energy $KE(T_{\rm{grain}})$ is randomly drawn from a
Maxwell-Boltzmann distribution at the grain temperature $T_{\rm{grain}}$.
So, the binding energy is:
\begin{eqnarray}
&&E_{\rm{Bind}}(i\to i-1) = \nonumber \\
&&\left[(i-1)E_{\rm{Bind,blk}} - \sigma S(i-1)\right] - \left[iE_{\rm{Bind,blk}} - \sigma S(i)\right] = \nonumber \\
&&-E_{\rm{Bind,blk}}-\sigma\left[S(i-1)-S(i)\right] = \nonumber \\
&&-E_{\rm{Bind,blk}} -
\sigma(36\pi)^{1/3}
\!\!\left(\frac{m_X}{\rho_X}\right)^{2/3} 
\!\!\!\!\!\left[(i-1)^{2/3}-i^{2/3}\right]
\end{eqnarray}
again for $S(i)=(36\pi)^{1/3} \left(\frac{m_Xi}{\rho_X}\right)^{2/3}$.

The rate of detachment of monomers from big grains is be given by
Eq.~\ref{eq:gi}. However, as previously mentioned, $g_i$ requires a
modification to correctly describe detachment from very small and/or
very cold clusters.  As Guhathakurta \& Draine (1989) realized,
clusters must have at least enough internal energy to supply a single
particle within the grain with at least the binding energy to eject
that particle from the cluster. However, even if the internal energy
of the cluster is above this threshold, we cannot be certain that this
energy will break the bond of a single monomer. In order to determine
the probability that this could happen, we must introduce quantization
of energy, modelling clusters as Einstein solids, such that $X_i$ has
$d_f=3i-6$ vibrational degrees of freedom. Assuming that these degrees
of freedom are harmonic oscillators with common frequency
$\omega_0=0.75kT_{Debye}$ and average energy $\hbar\omega_0$, and
given the internal energy $U$ of the cluster, we can determine the
total number of quanta in the cluster, $q=U/\hbar\omega_0$, as well as
the number of quanta required to eject a monomer,
$b=E_{\rm{Bind}}/\hbar\omega_0$. Using a combinatorial approach with
the preceding values, we calculate the probability that one degree of
freedom has \emph{at least} $b$ quanta. This (see Guhathakurta \&
Draine 1989) reduces the monomer detachment rate by:
\begin{equation}
C_q = \left(\frac{d_f}{q} + 1\right)^b \frac{q! (q-b+d_f-1)!}{(q+d_f-1)!(q-b)!}
\end{equation} 
where $\left(\frac{d_f}{q} + 1 \right)^b$ is the normalization factor
for $d_f \to \infty$.  Thus, the new detachment rate is
simply:
\begin{equation}
g_i^*= C_qg_i
\label{eq:gi3}
\end{equation} 
If we define $y\equiv q/d_f$ as the average number of quanta per
degree of freedom, $C_q$ is nearly unity when $y\gg b$ while $C_q\to
0$ as $y\to 0$ or $q\to b$; consequently, for $q<b$, $g_i^*=0$ and
monomers cannot be ejected from the cluster.
 
\subsection{Inert Carrier Gas Collision}
Even though we focus on homogeneous nucleation, we also consider
collisions with inert particles, such as hydrogen atoms, that are
inevitably mixed in the surrounding gas in any astrophysical
setting. We model this interaction as an event in which the particle
exchanges energy with the cluster and rebounds with a new velocity
distribution. The rate of inert collisions is as follows:

\begin{equation}
h_i=k_s n_{\rm{CG}} (36\pi)^{1/3} \left(\frac{m_X}{\rho_X}\right)^{2/3}
\sqrt{\frac{kT_{\rm{gas}}}{2\pi m_{\rm{CG}}}}\, i^{2/3}
\label{eq:hi}
\end{equation}
Although similar to $f_i$, this rate is dependent on the number
density of the inert gas, $n_{\rm{CG}}$ and on the mass of a single
gas particle. The change of internal energy for an inert collision is
simply the the change in kinetic energy after the collision:
\begin{equation}
\Delta U_{\rm{Col}} = KE(T_{\rm{gas}})-KE(T_{\rm{grain}})
\end{equation}
where both $KE(T_{\rm{gas}})$ and $KE(T_{\rm{grain}})$ are randomly
drawn from Maxwell-Boltzmann distributions at the gas temperature
$T_{\rm{gas}}$ and grain temperature $T_{\rm{grain}}$, respectively.
We make the approximation that as the inert gas particle comes in
contact with the cluster, it immediately attains the cluster's
temperature, and rebounds with that dependency. Even though this is
only an approximation, it ensures that over many collisions thermal
equilibrium is established.

\subsection{Photon Absorption}

The rate at which the cluster absorbs photons is derived from an
integrated form of the Planck law, given by:
\begin{equation}
  \alpha_i = 0.37(36\pi)^{1/3}
  C_{\rm{abs}}\left(\frac{m_X}{\rho_X}\right)^{2/3}
  \frac{\sigma_{SB}T_{\rm{rad}}^3}{k}i^{2/3}
\end{equation}
where $C_{\rm{abs}}$ is the correction term resulting from the fact
that clusters do not behave as blackbodies. This term is calculated
by:
\begin{equation}
C_{\rm{abs}} =  \left. \int \frac{\varepsilon(\nu,i)\nu^2
    \,d\nu}{e^{\frac{h\nu}{kT_{\rm{rad}}}}-1} \Bigg/ \int
  \frac{\nu^2 \,d\nu}{e^{\frac{h\nu}{kT_{\rm{rad}}}}-1} \right.
\end{equation}
where $\varepsilon(\nu , i) \leq 1$ is the emissivity of the material,
dependent on the size of the cluster, and the frequency of radiation
(Draine \& Lee 1984; Laor \& Draine 1993). For small grains, $C_{\rm{abs}}$
severely reduces the rate of photon absorption as the cluster only
significantly interacts with a few frequencies.
 
The change in internal energy for a photon absorption is given by:
\begin{equation}
\Delta U_{\rm{Abs}} = h\nu(T_{\rm{rad}})
\end{equation}
where $\nu$ is randomly drawn from the modified blackbody 
distribution at the radiation field temperature $T_{\rm{rad}}$, given by:

\begin{equation}
P(\nu,i) \propto \frac{\varepsilon(\nu,i)\nu^2}{e^{\frac{h\nu}{kT_{\rm{rad}}}}-1}
\end{equation}

\subsection{Photon Emission}

The rate at which the cluster emits photons is analogous to the rate
of photon absorption:
\begin{equation}
  \gamma_i = 0.37(36\pi)^{1/3}
  C_{\rm{emis}}\left(\frac{m_X}{\rho_X}\right)^{2/3}
  \frac{\sigma_{SB}T_{\rm{grain}}^3}{k}i^{2/3}
\end{equation}
except that it is evaluated at $T_{\rm{grain}}$ and $C_{\rm{emis}}$ is given by:
\begin{equation}
C_{\rm{emis}} = \int \frac{\varepsilon(\nu,i)\nu^2
  \,d\nu}{e^{\frac{h\nu}{kT_{\rm{grain}}}}-1}  
\Bigg/ \int \frac{\nu^2 \,d\nu}{e^{\frac{h\nu}{kT_{\rm{grain}}}}-1} 
\end{equation}
The change in internal energy for a photon emission is:
\begin{equation}
\Delta U_{\rm{Emis}} = -h\nu(T_{\rm{grain}})
\end{equation}
where $\nu$ is instead randomly drawn from the modified blackbody 
distribution at the grain temperature $T_{\rm{grain}}$:

\begin{equation}
P(\nu,i) \propto \frac{\varepsilon(\nu,i)\nu^2}{e^{\frac{h\nu}{kT_{\rm{grain}}}}-1}
\end{equation}

\subsection{Grain Temperature}

After each event, the internal energy of the cluster/grain is updated
according to the above equations. Once the internal energy is known,
the temperature of the cluster/grain can be calculated if the specific
heat of the material is known. The new temperature is then fed back into
the rate equations, a new event is computed and the cycle is repeated
until either the cluster evaporates or grows into a stable grain.

\section{Monte Carlo calculations}

\begin{figure}
\psfig{file=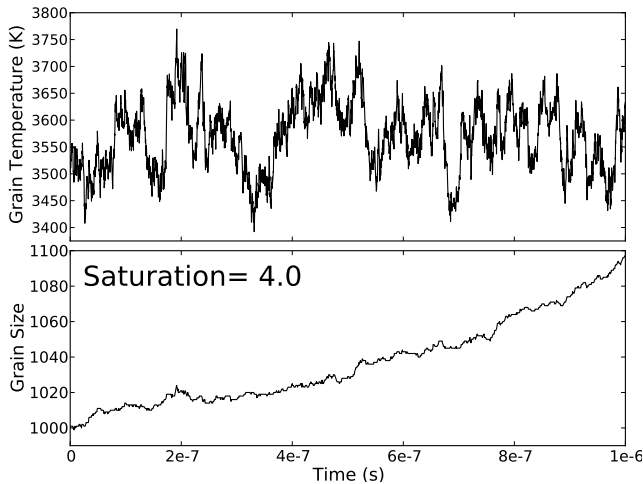,width=\columnwidth}
\caption{{Temperature and size evolution of a grain initially composed
   of 1000 carbon atom immersed in a hydrogen-carbon gas with
   $H/C=100$, $T_{\rm{gas}}=3500$~K, and with saturation $S=4$.}
\label{fig:thist1000}}
\end{figure}

 The physical process described above can be implemented in a Monte
Carlo computer code to follow the random oscillations of the grain
temperature and the grain fate under the various driving forces.  Here
and in the following, we apply the theory described above to clusters
of carbon atoms, precursors of carbonaceous dust grains. We adopt the
absorption coefficients for graphite grains from Draine \& Lee (1984)
and Laor \& Draine (1993), extrapolating their result to smaller
grains when necessary. The other graphite properties we adopt and
relevant references are summarized in Table~\ref{tab:graphite}.

An example of the thermal history of a relatively big carbonaceous
grain is shown in Figure~\ref{fig:thist1000}, where a grain of
initially 1000 carbon atoms is left free to evolve in a
hydrogen-carbon gas with number density ratio $H/C=100$, temperature
$T_{\rm{gas}}=3500$~K, and saturation $S=4$. The grain is observed to
grow as its temperature fluctuates with excursions as large as 200~K,
or $\sim5$ per cent of the grain temperature. Figure~\ref{fig:thist10}
shows instead the much more violent thermal history of a small grain,
injected in the same gas. In this case the temperature can fluctuate
by more than 100 per cent. Even under such violent fluctuations, the
grain eventually grows to a larger size.

\begin{table*}
\begin{tabular}{l|l}
  Property & Notes \\ \hline
  $\sigma=1500$ erg cm$^{-2}$ & Surface energy (Tabak et al. 1975)\\
  $\rho = 2.23$  g cm$^{-3}$ & Density of graphite \\
  $m_C=1.9944^{-23}$ g &  Mass of one carbon atom \\
  Eq. 3.3 of Guhathakurta \& Draine (1989) & Heat capacity  \\
  $T_{\rm{Debye}}=420$ K & Debye temperature (Guhathakurta \& Draine 1989)\\
  $n_{X,\rm{eq}}=\frac{6.9*10^{13}e^{\frac{-844282}{T_{\rm{grain}}}}}{kT_{\rm{grain}}}$
  cm$^{-3}$ &  Equilibrium gas density \\
  $E_{\rm{Bind,blk}}=1.1831760^{-11}$ erg & Bulk binding energy of graphite
\end{tabular}
\caption{{Graphite properties used in this paper.}
\label{tab:graphite}}
\end{table*}

\subsection{Grain Stability} 

\begin{figure}
\psfig{file=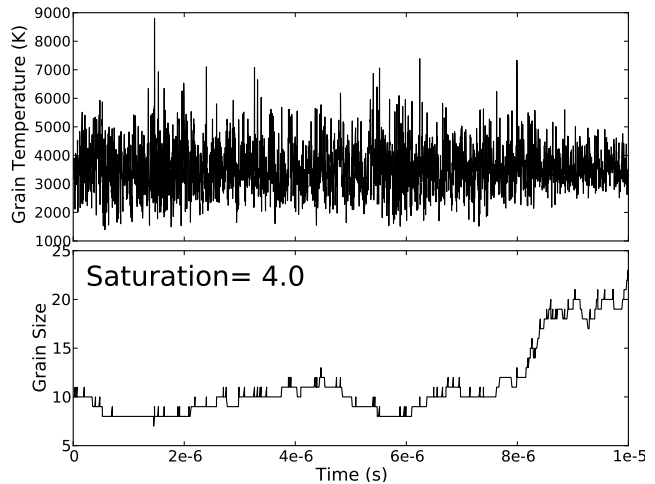,width=\columnwidth}
\caption{{Same as Fig.~\ref{fig:thist1000} but for an initially
    smaller grain made by 10 carbon atoms.}
\label{fig:thist10}}
\end{figure}

An important consequence of our analysis of the fluctuations of the
grain temperature under the effect of quantized energy exchanges with
the gas and the radiation field is that the grain stability becomes a
statistical concept instead of an exact concept. In the classical
approach, when the Gibbs free energy is maximized, the grain is
supposed to be in an unstable but exact equilibrium, able to remain at
the stable size forever. Any cluster smaller than the critical size is
doomed to evaporate and any grain bigger than the critical size will
eventually grow. In our approach, two grains of the same size can be
in different stability conditions (one stable and the other unstable)
if they are at different temperatures. Moreover, a grain that is
stable can become an unstable cluster without changing its size by
absorbing a photon that raises its temperature and vice versa. A
critical size that depends only on the gas temperature and saturation
- reminiscent of the one of the classical theory - can be defined as
the size for which the probability of eventual evaporation is the
same as the probability of eventual growth into a stable grain.

Here we discuss stability by finding the saturation at which a certain
cluster becomes stable, according to the above
definition. Unfortunately the photon absorption interactions become so
numerous at low temperatures ($T_{\rm{gas}}<2500$~K) that running the
code until a statistically sound number of C atom attachments or
detachments is observed becomes practically unfeasible. For that
reason, in the following we concentrate on the high temperature
nucleation.

Classically, the saturation at which a cluster with size
$i$ is stable, or critical, is given by:

\begin{equation}
S=e^{\left(\frac{32\pi}{3}\right)^{1/3}
  \left(\frac{m_X}{\rho_X}\right)^{2/3} \frac{\sigma}{kT_{\rm{grain}}}
  i^{-1/3}}
\end{equation}

\begin{figure}
\psfig{file=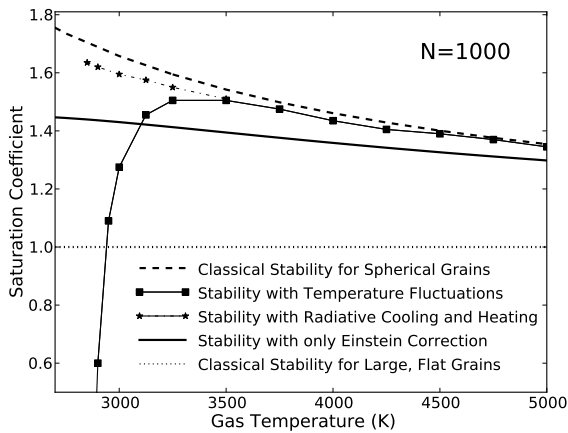,width=\columnwidth}
\caption{{Stability plot in the $T_{\rm{grain}}-S$ plane for a
    carbonaceous cluster with $i=1000$ carbon atoms immersed in a
    hydrogen-carbon gas with $H/C=100$. Both the solid line with
    square markers and the dash-dot line with star markers, show the
    results of this paper.  Actual calculations were performed only at
    the marked locations. Both lines show stability with temperature
    fluctuations, however, the solid line does not include an external
    radiation field, while the dash-dot line does, incorporating the
    heating effect of photons. The radiation field is assumed to be a
    black body at the temperature of the gas. For higher temperatures,
    radiation effects become negligible and the lines converge. The
    dashed line shows the stability for spherical clusters in the
    classical approximation in which the temperature of the cluster is
    constant and equal to the temperature of the gas. The thick solid
    line shows instead the stability line if the Einstein model is
    applied to the standard theory but temperature fluctuations are
    neglected. }
\label{fig:stabil1000}}
\end{figure}

\noindent
Since $T_{\rm{grain}}$ is not constant in our approach, we determine
the stability saturation for a given cluster at a specific gas
temperature by proceeding iteratively as follows.  We start from a
fairly large saturation (e.g. $S=10$) and perform a minimum of 10
simulations in which the grain may evolve at that saturation.  Each
simulation is halted when the initial cluster has changed its size by
at least 10 per cent\footnote{For large clusters ($i\ge100$) this is
  more than enough to determine the final fate of the
  cluster. However, for small grains we halt the simulations after a
  70 per cent change of the cluster initial size.}.  If all or most of
the simulations resulted in grain growth, the estimated saturation was
too large, and if all or most of the simulations resulted in grain
evaporation, the estimated saturation was too small. We adjust our
estimate of the critical cluster size and repeat the process,
narrowing in on the saturation for which half of the clusters grow and
half evaporate. In this way we can obtain a new stability curve, i.e.,
the location in the $T_{\rm{gas}}-S$ plane where a grain of a given
size has the same probability of growing or evaporating. This is
analogous to the critical cluster size in the classical nucleation
theory. Thus, if the gas saturation is greater than the stable
saturation the grain is more likely to grow.

\begin{figure}
\psfig{file=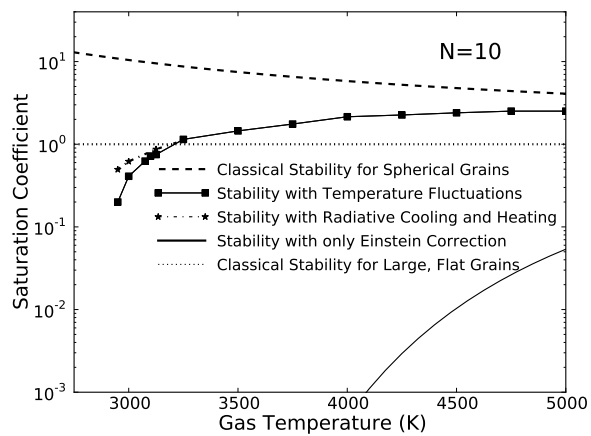,width=\columnwidth}
\caption{{Same as Figure~\ref{fig:stabil1000} but for a smaller
    cluster with $i=10$ carbon atoms. Even with the effect of an
    external radiation field (stars) the Einstein correction
    dominates.}
\label{fig:stabil10}}
\end{figure}

The stability plot for clusters with $i=1000$ is shown in
Figure~\ref{fig:stabil1000}. For all gas temperatures, the new
stability curve is lower than in the standard theory, generally
allowing lower saturations in which clusters can grow.  This is a
result of three competing effects. At high temperatures
($T_{\rm{gas}}>3000$~K) the strongest effect is the Einstein model
correction that makes the clusters much more resilient to
evaporation. At lower temperatures, the grains are made more stable
(even more than the Einstein correction prediction) by the fact that
their temperature is lower than the gas temperature (Lazzati 2008).
Both these stabilizing effects are balanced by the effect of
temperature fluctuations that tend to make clusters unstable. The drop
of the stability line at low temperatures is expected to be reduced in
the situation in which the radiation field is dense, so that
interactions mediated by photons can also have a heating effect, as
well as a cooling effect. In the case of a blackbody radiation field
in thermal equilibrium with the gas, the drop is absent, as shown by
the star symbols in Figure~\ref{fig:stabil1000}. This last condition
may be relevant for dust production in a supernova explosions.

Figure~\ref{fig:stabil10} shows instead the stability line for smaller
clusters with $i=10$ carbon atoms. In this case, the Einstein
correction term has a big influence, allowing its effect to dominate
over the entire temperature range that we investigated.  Even the
heating effect of a radiation field in thermal equilibrium with the
gas does little to diminish the effect of the Einstein model
correction (see the star symbols in Figure~\ref{fig:stabil10}. This
steep drop makes determining stability for lower temperature more
difficult. Also, smaller clusters are more susceptible to random
fluctuations, inherent in our code, requiring more simulations, at
least a minimum of 30, to determine the new stability curve.

\begin{figure}
\psfig{file=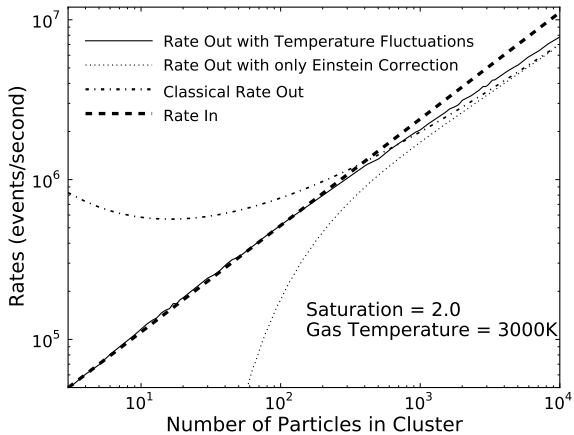,width=\columnwidth}
\caption{{Attachment and detachment rates from carbonaceous clusters
    for $S=2$ saturation and gas temperature $T_{\rm{gas}}=3000$~K.}
\label{fig:ratesinout}}
\end{figure}

Another way to look at cluster stability and critical cluster size is
to plot the attachment and detachment rates as a function of cluster
size for given conditions of gas temperature and saturation. With the
use of a Monte Carlo code like ours it is not straightforward to
compute the average detachment rate for a given cluster size. That is
because in a run, the cluster size varies (see
Figure~\ref{fig:thist1000} and~\ref{fig:thist10}). In order to compute
the average detachment rate from a cluster of a given size $i$ we run
the code artificially maintaining the cluster size constant. When an
attachment happens, we compute the temperature that the $i+1$ cluster
would attain, but we keep the size equal to $i$. Analogously, when a
detachment happens, we compute the temperature that the $i-1$ cluster
would have but we maintain the size constant to $i$. This artificial
forcing is equivalent to assuming that the temperature distribution of
the size $i$ cluster is not different from the temperature
distribution of the $i-1$ and $i+1$ clusters.

Figure~\ref{fig:ratesinout} shows attachment and detachment rates from
clusters containing between $i=3$ and $i=10000$ carbon atoms embedded
in a saturation $S=2$ hydrogen-carbon gas with $H/C=100$ and
temperature $T_{\rm{gas}}=3000$~K. The attachment rate is shown as a
thick dashed line while detachment rates for different assumptions are
overlaid. The classical detachment rate is shown with a dash-dot line,
yielding a critical cluster of $i_{\rm{C}}=350$ carbon atoms. The
dotted line shows the predicted rate if the Einstein model is
considered but thermal fluctuations are neglected. The graph clearly
shows that by itself the Einstein correction is too strong, making
even the smallest clusters stable at a modest saturation. This would
result in a catastrophic nucleation of very small grains. The
detachment rate with thermal fluctuations included is shown with a
black line. Interestingly, the attachment and detachment rates almost
trace each other for the small, unstable clusters. This results from
the fact that the biggest thermal fluctuations are due to the
attachment-detachment process, since the binding energy is typically
much larger than the photon and collision energies. As a consequence,
an attachment results in a sharp increase in the cluster temperature
that increases the detachment rate by a large factor, increasing the
probability of a detachment, which in turn cools the cluster
significantly.  This locking of the attachment and detachment events
is particularly evident for pure gases and for $k_s=1$, in which every
collision of the cluster is with a monomer that attaches to the
cluster. Less pure gases show this tracking to a lesser extent
(typically only for small clusters). In terms of the critical cluster
size, we see that the critical cluster is smaller than the one
predicted classically, but not by a huge factor. The complete theory
with fluctuations predicts a critical cluster of approximately
$i_{\rm{C}}=100$ carbon atoms.

\subsection{Nucleation Rates}

\begin{figure}
\psfig{file=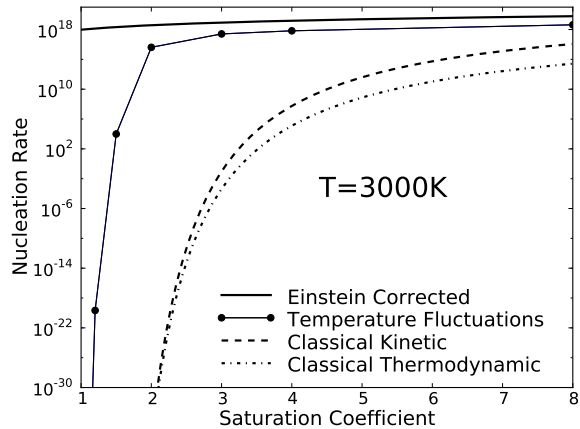,width=\columnwidth}
\caption{{Nucleation rate as a function of saturation for carbonaceous
    grains from a hydrogen-carbon gas with $H/C=100$ and temperature
    $T_{\rm{gas}}=3000$~K.}
\label{fig:j3000}}
\end{figure}

Once the average rates out for all sizes between $i=3$ and
approximately twice the critical size have been computed, the
nucleation rate can be calculated with a modification of
Eq.~\ref{eq:j}:
\begin{equation}
  J=\frac{n_X}{2} f_2 \left[ 1+\sum_{j=3}^{2i_{\rm{C}}} \left(
      \prod_{k=3}^j\frac{g_i}{f_i}\right)\right]^{-1}
\label{eq:jnew}
\end{equation}
Since the detachment rate is always significantly less than the
attachment rate for $i=2$, due either to linearity of the molecular
bond (thus a dramatic change in the internal energy) or from a drastic
drop associated with the Einstein correction term, we realize that we
cannot clearly define a nucleation rate directly from the gas
phase. Instead, we assume that all of the gas particles have formed
$i=2$ clusters, and compute $J$ using half of the number density of
the gas. The extreme robustness of the $i=2$ clusters is probably due
to the failure of the capillary approximation for very small clusters,
as discussed below.

\begin{figure}
\psfig{file=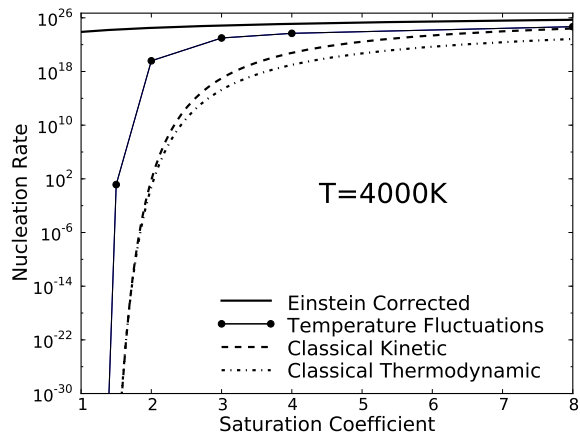,width=\columnwidth}
\caption{{Same as Figure~\ref{fig:j3000} but for a gas at temperature
    $T_{\rm{gas}}=4000$~K.}
\label{fig:j4000}}
\end{figure}

\begin{figure}
\psfig{file=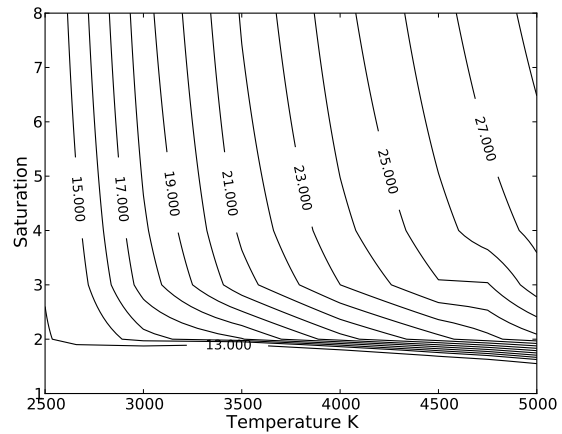,width=\columnwidth}
\caption{{The log of the nucleation rate for carbonaceous grains from a
    hydrogen-carbon gas with $H/C=100$.}
\label{fig:j}}
\end{figure}

Figure~\ref{fig:j3000} shows the rates computed with our Monte Carlo
code compared with rates obtained by other methods. The dash-dot line
shows the rate obtained from the thermodynamic approach (e.g. Kozasa
et al. 1989, 1991; Todini \& Ferrara 2001; Nozawa et al. 2003, 2008,
2010), while the dashed line shows the rate obtained from the kinetic
theory under the assumption of thermal equilibrium between the cluster
and the gas. A solid line shows the nucleation rate if the Einstein
model correction is taken into account, but thermal fluctuations are
neglected. As expected, this rate is very high since the correction
makes clusters of all sizes stable. Finally, a solid line with circles
shows the result with temperature fluctuations. Since, as we saw
before, the combined effect is that of making the clusters more
stable, the nucleation rate is significantly larger than the classical
result, especially at moderate saturation. Figure~\ref{fig:j4000}
shows the same results for a hotter gas with
$T_{\rm{gas}}=4000$~K. Finally, Figure~\ref{fig:j} shows a contour
plot of the nucleation rate as a function of both gas temperature and
saturation. It appears that the nucleation rate is highly dependent on
temperature at high saturations, but becomes increasingly dependent on
saturation as the threshold $S=2$ approaches.

\begin{figure}
\psfig{file=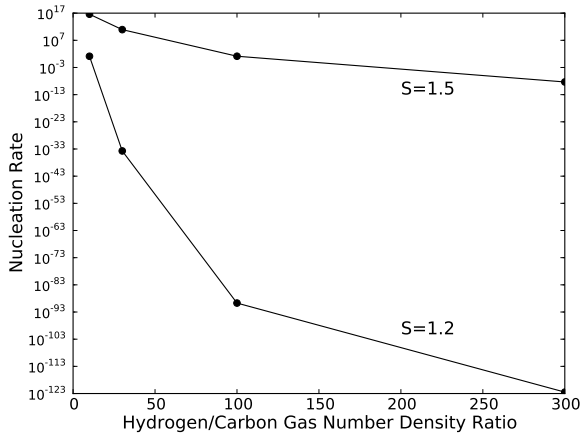,width=\columnwidth}
\caption{{Nucleation rate as a function of the carbon concentration in
    the gas phase for a hydrogen-carbon gas at saturations 1.2 and 1.5
    and temperature 4000 K. For low saturations, higher H/C ratios
    drastically lower nucleation rates.}
\label{fig:hc1}}
\end{figure}

\begin{figure}
\psfig{file=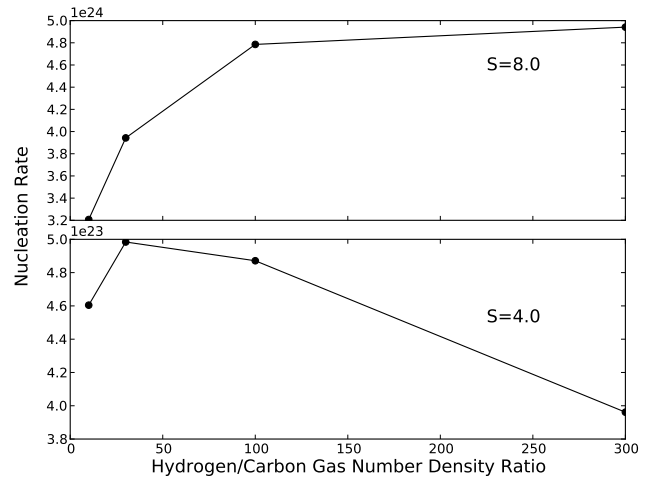,width=\columnwidth}
\caption{{Nucleation rate as a function of the carbon concentration in
    the gas phase for a hydrogen-carbon gas at saturations 4.0 and 8.0
    for 4000 K.  There appears to be a transition such that higher
    saturations instead allow higher H/C ratios to increase nucleation
    rates. }
\label{fig:hc2}}
\end{figure}

A final, important result, if not the most important result of this
work, is shown in Figures~\ref{fig:hc1} and~\ref{fig:hc2}.  The
figures shows the nucleation rate of carbonaceous grains as a function
of the carbon concentration in the gas, for $10<H/C<300$ and for
various saturation values. According to the classical theory, both
thermodynamic and kinetic, the nucleation rate should only depend on
the partial pressure (or partial density) of the compound that
participates in the nucleation, carbon in our case. Instead, we see
that if thermal fluctuations are relevant, the nucleation rate depends
quite strongly on the concentration. The dependence of nucleation rate
on concentration is complex and not straightforward to interpret. For
low saturation, the nucleation rate decreases as the relative density
of the carrier gas increases, i.e., a pure carbon gas nucleates much
faster than a mixed gas (Figure~\ref{fig:hc1}). At intermediate
saturations, the fastest nucleation is observed at intermediate
concentrations ($H/C\sim50$ for $S=4$, lower panel of
Figure~\ref{fig:hc2}). Finally at high saturations, the nucleation
rate increases at low concentrations (or high $H/C$ values). The trend
seen in Figure~\ref{fig:hc1} at low saturations is easy to
explain. The presence of carrier gas particles creates a thermal
contact between the cluster and the gas that established thermal
equilibrium without requiring growth or evaporation, breaking the
strict growth$\to$heating$\to$evaporation$\to$cooling$\to$growth
behavior outlined above and making the clusters more vulnerable. The
trends observed in Figure~\ref{fig:hc2} are instead more difficult to
fully understand. Inspection of the rate plots (analogous to
Figure~\ref{fig:ratesinout}) suggests that the increase of the
nucleation rate is due to the behavior of the grains just above the
critical size rather than to the attachment and detachment rates from
sub-critical clusters.

\section{Discussion}

We have presented a model for the physics of the nucleation rate of
dust grains from the gaseous phase that includes the role of thermal
fluctuations of the clusters as well as corrections for the detachment
rates from nanoscale clusters. Our implementation is based on the
kinetic theory of nucleation (Becker \& D\"oring 1935, Kashchiev
2000), but incorporates some fundamental changes in the way in which
the thermal balance of the forming clusters is considered. First, we
included a modification to the detachment rates from small grains,
following the Einstein model developed by Guhathakurta \& Draine
(1989). This is a stabilization effect that takes into account the
fact that very small grains at low temperature need to use a large
fraction of their total internal energy to brake a bond and eject a
monomer. As a consequence, the ejection rate does not scale linearly
with the surface but becomes very small for small grains. In addition,
we consider the fluctuations of the grain temperature as the grain
exchanges energy with the gas and the radiation field. Since all the
energy exchanges are quantized (through photons or collisions), the
temperature of the clusters does not attain an equilibrium value but
rather continuously fluctuates around it. We find that the combination
of these two effects results in an increased tendency of the clusters
to grow, and as a consequence, yields higher nucleation rates. We also
find that the nucleation rate depends on the concentration of the
nucleating compound in the inert carrier gas, with higher rates for
higher concentrations, especially at moderate saturation. This is a
new result that has not been predicted by any other theory and can
potentially become a crucial test for this model.

Our study has, like most nucleation studies, some important
limitations that deserve some discussion and could sizably affect the
numerical values of the critical clusters and nucleation rates.

\begin{itemize}

\item First, we were able to numerically implement the theory only for
  high temperatures, for which the rate of particle interactions
  (cluster-monomer or cluster-carrier gas) outnumber the radiation
  interactions (carrier-photon). We cannot therefore give any result
  for the interesting cold temperature regime, where supercooling of
  the cluster can happen, as discussed in Lazzati (2008). A possible
  solution for the low temperature regime is to treat radiation
  interactions as continuous and concentrate on the quantized nature
  of the particle interactions only. Such implementation is under
  study.

\item More fundamentally, we have been implicitly adopting the
  capillary approximation, i.e., we have assumed that the surface
  energy of graphite clusters is the same of the one of big graphite
  chunks and we have assumed that the concept itself of surface energy
  applies to clusters as small as a handful of carbon atoms, for which
  a surface is a hard concept to define. It is likely that this
  results in an overestimate of the binding energies of small
  clusters, ultimately resulting in an overestimate of the nucleation rate.

\item Another potential overestimating factor is that we assumed the
  sticking coefficient to be $k_s=1$, i.e., all incoming monomers that
  collide with the cluster stick to it and make the cluster grow. The
  nucleation rate is directly proportional to the sticking coefficient
  and, since $k_s\le1$, our nucleation rates are overestimated if the
  real coefficient is different from 1. In addition, reality is likely
  more complex. $k_s$ is probably dependent on the number of atoms in
  the cluster. Clusters that fill all the bonds of carbon atoms (such
  as fullerenes, for example), are likely to have very small values of
  $k_s$ since it is hard to make them grow beyond their closed
  structure. An $i$-dependent value of $k_s$ would, again, result in a
  decrease of the nucleation rate.

\item We have neglected the possibility of cluster-cluster
  interactions. Even though these should not be numerous, they can be
  relevant for a particularly pure gas of the nucleating compound.

\item We have assumed that even the smallest clusters emit continuum
  radiation. It is instead likely that they will emit radiation in
  bands becoming more and more narrow as the size decreases.

\item We have assumed that all the energy released as a bonding of a
  new monomer happens is transformed into internal energy of the
  cluster/grain. However, it is possible that some of that energy is
  carried away by a photon. If this were true, smaller temperature
  fluctuations would be seen in consequence of attachments of new
  monomers.

\end{itemize}

Despite all these limitations, we believe the two main conclusions of
this work hold. First, thermal fluctuations are important in the
nucleation process. Second, if this is true, the nucleation rate
should depend on the concentration of the nucleating compound. The
difference of the nucleation rates resulting from different theories
should warn us that a lot has still to be learned in this fascinating
field and conclusions based on the blind application of classical
nucleation rates should be taken with care.

\end{document}